# DUST PARTICLE SIZE, SHAPE AND OPTICAL DEPTH DURING THE 2018/MY34 MARTIAN GLOBAL DUST STORM RETRIEVED BY MSL CURIOSITY ROVER NAVIGATION CAMERAS

H. Chen-Chen*, S. Pérez-Hoyos, and A. Sánchez-Lavega.

*Departamento de Física Aplicada I, Escuela de Ingeniería de Bilbao, Universidad del País Vasco (UPV/EHU). Bilbao 48013, Spain*

* To whom correspondence should be addressed at: **hao.chen@ehu.eus**

**Abstract**

Martian planet-encircling dust storms or global dust storms (GDS), resulting from the combined influence of local and regional storms, are uncommon aperiodic phenomena: with an average frequency of approximately one every 3-4 MY, they produce a substantial rise in the atmospheric dust loading that lasts from weeks to months and have a significant impact on the atmospheric properties, energy budget, and global circulation. During the 2018/MY34 global dust storm, initiated at $L_S = 185º$ (30-31 May 2018), an intensive atmospheric science campaign was carried out by the Mars Science Laboratory (MSL) rover to monitor the environmental parameters at Gale Crater. We contribute to previous studies with independent retrievals to constrain the dust opacity and characterise the aerosol particle properties, including: size, shape and single scattering phase function. An iterative radiative transfer retrieval procedure was implemented to determine the aerosol parameters that best fit the angular distribution of sky radiance at forward and backward scattering regions observed by MSL Navigation Cameras (Navcams) during the 2018/MY34 GDS. The MOPSMAP aerosol database and Double Henyey-Greenstein (DHG) analytical single scattering phase functions were used to model the Martian dust aerosol. Outcomes of this study show a steep rise in dust opacity from pre-storm levels of 1.2 up to $\tau > 9$, correlated to particle size variations from 1 to 4 μm. DHG phase functions are characterised with an average asymmetry parameter of $g = 0.60\pm0.11$ during the storm, diverging from values of around $0.71\pm0.06$ for the same period in previous MY. Best fitting simulations to backscatter observations for high-opacity periods were generated by a mixture of spheroids following a log-normal distribution of aspect ratios centred on $2.8\pm0.9$, in contrast to values of 1.8 in post-storm sols, thus pointing to more irregular particle shapes at the peak of the dust storm.

## 1. Introduction

Airborne dust particles play a predominant role in the variability of the Martian atmosphere: by absorbing and scattering the incoming solar radiation, they have a direct impact on its thermal structure and dynamics (Gierasch and Goody, 1972; Pollack et al., 1979). Dust is always present in the atmosphere, its abundance varies significantly with season and from year to year, and its activity covers a wide range of scales, from dust devils to local, regional and global dust storms (e.g., Martin and Zurek, 1993; Lemmon et al., 2015; Wang and Richardson, 2015).

Planet-encircling dust storms or global dust storms (GDS) are the most dramatic dust events since they blur surface albedo markings as observed before the spacecraft exploration era (Martin and Zurek, 1993; Strausberg et al., 2005; Cantor, 2007; Kahre et al., 2017 and references therein). Martian global dust storms result from the rapid expansion and combined influence of multiple local and regional storms (e.g., Zurek and Martin, 1993). They produce substantial dust optical depths ($\tau > 3$) in the visible band over large portions of the planet, lasting for weeks to months, and have significant effects on diurnal pressure variations, atmospheric heating, and global circulation (Newman et al., 2002, Zurek and Martin, 1993; Strausberg et al., 2005). While local and regional dust storms are common every Martian Year (MY, following convention by Clancy et al., 2000) during the northern hemisphere autumn and winter seasons ($L_S = 180º$ to $360º$), and with preferred locations (e.g., Wang and Richardson, 2015), planet-encircling dust events are unpredictable aperiodic phenomena, with an average frequency of approximately one every 3-4 MY (e.g. Zurek and Martin, 1993; Smith, 2008; Wolkenberg et al., 2020). Six confirmed GDS have been observed by instruments either in Mars orbit or on its surface (for recent reviews, see Kahre et al. 2017 and references therein): 1971/MY9 by Mariner 9 (e.g., Martin, 1974; Toon et al., 1977), two dust storms in 1977/MY12 by the Viking mission (e.g., Briggs et al., 1979; Pollack et al., 1979), 2001/MY25 by the Mars Global Surveyor (MGS) (Smith et al., 2002; Clancy et al., 2003; Cantor, 2007), 2007/MY28 by Mars Odyssey (Wang and Richardson, 2015), Mars Express (Vincendon et al., 2009), and the Mars Exploration

Rovers (MER) *Spirit* and *Opportunity* (Lemmon et al., 2015), and in 2018/MY34 by six spacecraft in orbit and two rovers on the surface (e.g., Montabone et al., 2020; Smith, 2019; Kleinböhl et al., 2020, Guzewich et al., 2019; Lemmon et al., 2019, and other publications in the same special issue). However, there are limited systematic studies of the atmospheric dust loading and particle properties during global dust events using surface-based observations to complement orbital observations (e.g., Dluglach et al., 2003, Lemmon et al. 2015; 2019).

In 1977/MY12, Viking landers (VL1, VL2) atmospheric visible 670-nm opacity measurements at the arrival of the first storm ($L_S \sim 205°$) showed values of about 3.0 and 2.0 at VL1 and VL2 sites, respectively; while for the second storm ($L_S \sim 275°$) abrupt increases were retrieved with opacities ranging from 3.7 up to 9 for VL1, and > 3 for VL2 (Pollack et al., 1977; 1979). Pollack et al. (1995) derived the particle size distribution during non-dusty and storm periods, resulting in effective radius values of 1.85 and 1.52 µm, respectively, with $v_{eff}$ fixed to 0.5. The eddy mixing and renewal of dust by local storms were invoked as possible explanations for the lack of size evolution (Toon et al., 1977; Murphy et al., 1990; Pollack et al., 1995). Particle shapes were reported as non-spherical, but rather plate-like (Pollack et al., 1979; 1995); and the dynamical modelling of the decay phase (Murphy et al., 1990) concluded that particles should be disc-shaped in order to match the opacity decay rates observed by the Viking landers. During the 2007/MY28 event at $L_S \sim 277°$, the *Spirit* rover measured 880-nm optical depth values of up to 4.3; while peaks of 4.6 to 5 were derived by *Opportunity* (Lemmon et al., 2015). MER Navcam particle size retrievals by Smith and Wolff (2014) showed a trend for larger aerosol particles during the active phase of a dust storm, with sizes of ~2.0 µm for opacity values of 3.0; also confirmed by MER Pancam opacity 440-nm to 880-nm colour ratios (Lemmon et al., 2015). Phase functions retrieved by Pancam low-Sun sky survey showed that Mie theory for spherical particles fails to provide good matches at large scattering angles (Smith and Wolff, 2014).

The 2018/MY34 global dust storm initiated at $L_S = 184.9°$ (30-31 May 2018, MSL sol 2067) with precursor storms expanding and merging in the Acidalia and Utopia Planitia regions (30-60° N). See for example: Sánchez-Lavega et al. (2019), Guzewich et al. (2019), Wolkenberg et al. (2020), and further references in the special issue for a chronology, description and analysis of the onset, expansion, evolution and decay phase of the phenomenon using both surface-based and in-orbit observations. An intensive atmospheric science campaign was carried out by the MSL *Curiosity* rover at Gale Crater for monitoring the storm, starting on sol 2075 ($L_S = 188.7°$, 7 June 2018) and finishing on sol 2169 ($L_S = 248.2°$, 11 September 2018) (Viúdez-Moreiras et al., 2019; Smith et al., 2019; Guzewich et al., 2019). During this planet-encircling dust event, MSL Mastcam direct Sun-imaging measurements of column dust opacity showed a steep increase from 0.6 up to 8.5 over a period of about 10 sols (Lemmon et al., 2019). Dust particle size retrievals using cross-sky and near-Sun imaging surveys by Mastcam, ChemCam sky spectral data, and ultraviolet flux measurements by REMS showed an abrupt increase from 1.5 to > 4 µm in effective radius, with increasing opacity (Lemmon et al., 2019).

The MSL rover is equipped with a set of 4 Navigation Cameras (Navcam) located at the rover's remote sensing mast to support the operation of the vehicle during its drive across the surface (Maki et al., 2012). Although not designed as a scientific instrument, Navcam images can be used as an alternative source of data for atmospheric studies and characterising Martian dust aerosol properties (Soderblom et al., 2008; Smith and Wolff, 2014; Moores et al., 2015; Wolfe and Lemmon, 2015; Chen-Chen et al., 2019a, 2019b; Smith et al., 2019).

In this work, we study the evolution of dust aerosol particle properties during the 2018/MY34 global dust storm. We contribute to previous studies with independent retrievals using MSL Navcam observations. The systematic and continuous characterisation of the behaviour of dust single scattering phase function and particle's shape performed in this present work provides aerosol modelling inputs to be used in regional and global circulation model simulations, and to gain insight into the interrelationship between particle properties and dust cycle dynamical processes, including its transport and deposition rates.

This manuscript is structured as follows. In Section 2 we present the Navcam image sequences used in this work. Section 3 describes the methodology followed, and in Section 4 we show and discuss the results of this study; including some sensitivity testing of the different parameters and a comparison with retrievals from other instruments. Finally, Section 5 summarises the main findings of this study and lays out future research prospects.

## 2. Observations

MSL Navcam imagers are build-to-print copies of Mars Exploration Rover (MER) mission cameras: they have a 45º field-of-view and are equipped with a 1024x1024 pixel CCD detector, and a broadband visible filter of range 600-800 nm (Maki et al., 2003; 2012). During the 2018/MY34 global dust storm, regular observations were retrieved by MSL Navcams under multiple viewing geometries. We take advantage of Navcam imagers' capacities and their flexible pointing (Soderblom et al., 2008; Maki et al., 2012) to retrieve the angular variation of the sky radiance at multiple scattering geometries.

For this work, Sun-pointing images and observation sequences pointing at the anti-solar point, i.e. the backward scattering region, were selected from the accumulated image data-set: the decay of the sky radiance up to about 40º away from the solar disc centre is highly sensitive to the size of the particle; whereas the angular distribution of sky radiance at intermediate and large scattering angles provides useful information of the particle shape-dependent parameters (e.g., Hansen and Travis, 1974; Kaufman et al., 1994; Pollack et al., 1995).

We used the non-processed binary data record produced by the instrument, i.e., Engineering Data Record or EDR (Alexander and Deen, 2018). Radiometric calibration and geometric reduction of Navcam raw EDR files were performed following the procedure described in Chen-Chen et al. (2019a). The list of Navcam observations used in this study is provided in Table A1 of the appendix to this manuscript.

2.1 Sun-pointing sequences

Two image sets of Navcam Sun-pointing observations were used in this study: the Surface Attitude Pointing and Positioning (SAPP) sequence, commanded for the calculation of the rover's attitude (Maki et al., 2003); and the NCAM00600 sequence. Examples of Navcam Sun-pointing images are shown in Figure 1. SAPP observations are regularly programmed during the MSL mission. In these images, the solar disc is fully contained within the detector frame and they are obtained between 13h and 16h LTST, when the solar elevation angle ranges from approximately 65º to 30º. Navcam observations labelled with "NCAM00600" were retrieved during the GDS in a regular manner, approximately 1 sequence every 1 or 2 sols. The sequence for each sol consist of 3 observations: 1 near-Sun observation, in which the solar disc is outside the image; and 2 observations with the solar disc totally contained within the image frame, similar to the SAPP set, with different exposure times. For this study, only the latter two images were considered, as the precise location of the solar disc centre is required for an accurate geometric reduction and calculation of the scattering angles.

2.2 Backscatter observations

A total of 14 sequences of Navcam "NCAM00565" backscatter observations were retrieved during the 2018/MY34 GDS. The purpose of these sequences is to monitor the status of the imager and retrieve sky flats (Alexander and Deen, 2018). The observations were performed around 17h LTST, when the solar elevation angle is between 15º and 20º; with cameras pointing at the backward scattering or anti-solar region. Each of these sequences comprises one full-frame observation (EDRF, 1024x1024 pixels) and four rover onboard downsampled images (EDRD, 128x128 pixels). The brightness as a function of the scattering angle, usually ranging from 90º to 160º, was then retrieved by sampling along the solar principal plane.

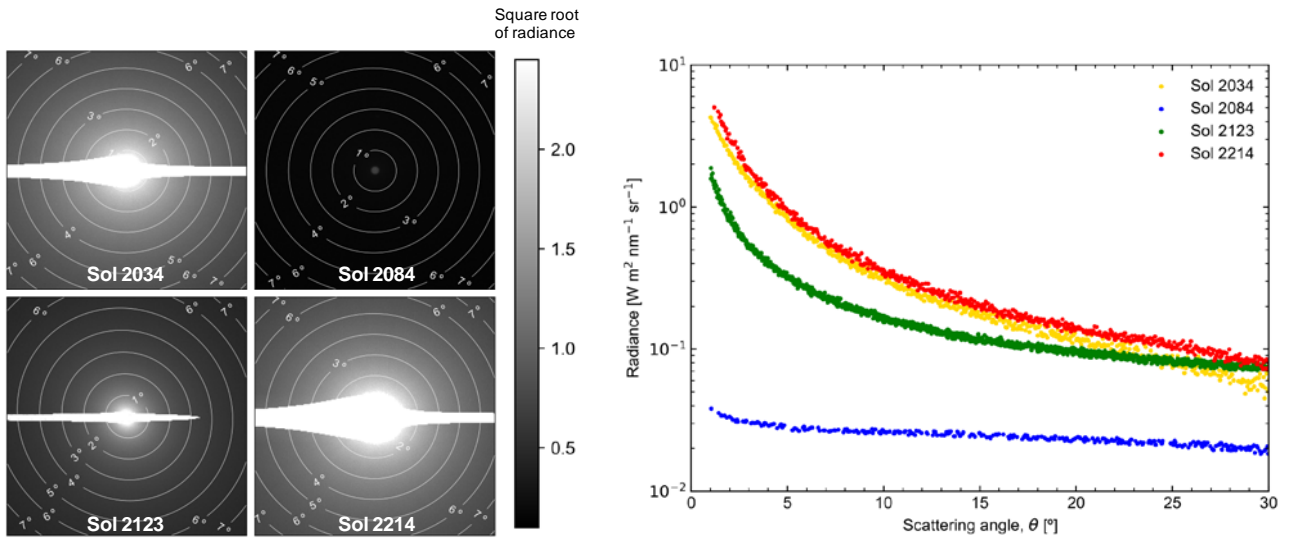

*Fig. 1. Navcam Sun-pointing observations during 2018/MY34 GDS. Left: Navcam calibrated observations for sols 2034 (pre-GDS), 2084 (close to GDS peak), 2123 (decay phase) and 2214 (post-GDS). Radiometric correction and geometric reduction were performed on the EDR files to convert the pixel DN counts into absolute radiance (W m$^2$ nm$^{-1}$ sr$^{-1}$), and to calculate the scattering angle. For clarity, the square root of the corresponding radiance value is plotted. On sol 2084, the solar disc can be seen as the high dust opacity ($\tau \sim 7.0$, Lemmon et al., 2019) prevents image's pixels from saturation. Right: the sky radiance as a function of scattering angle sampled along a diagonal path, starting at the centre of the solar disc and finishing on the upper-left corner of the image (e.g., Soderblom et al., 2008; Chen-Chen et al., 2019a).*

## 3. Methodology

In this study, the angular distribution of sky brightness observed by MSL Navcam was compared with radiative transfer simulations to retrieve the aerosol properties during the 2018/MY34 global dust storm. The multiple scattering of incoming solar radiation by dust particles into the line-of-sight of the cameras is modelled using the radiative transfer forward model described in Chen-Chen et al. (2019a). It uses the discrete ordinates method (Stamnes et al., 1988) to simulate the observed sky brightness for an upward-looking geometry as a function of the column dust opacity and various aerosol parameters. Except stated otherwise, we use in this work the same overall description of the atmosphere as presented in Chen-Chen et al. (2019a, 2019b).

The radiative transfer code requires only 3 parameters at each layer of the discretised atmosphere for characterising the airborne aerosol: the single scattering albedo ($\omega_0$), the single scattering phase function $P(\theta)$, and the optical depth at the specific layer. In this study, we have considered two different approaches for defining these aerosol parameters: the MOPSMAP aerosol database (Gasteiger and Wiegner, 2018) and Double Henyey-Greenstein phase functions (Kattawar, 1975).

### 3.1 MOPSMAP tool
The MOPSMAP package is a tool for modelling optical properties of aerosol particles: it consists of a dataset of pre-calculated optical properties for single aerosol particles, and a Fortran routine that calculates and returns the properties of user-defined aerosol ensembles from this dataset. Depending on the defined particle characteristics, different approaches are used for retrieving the optical parameters including: Mie theory, T-matrix, improved geometric optics, and the discrete dipole approximation. For further details please refer to Gasteiger and Wiegner (2018) and references therein.

For modelling the Martian dust aerosol in Navcam's effective wavelength (650 nm), in this study we have used the complex refractive indices reported in Wolff et al. (2009), and a log-normal particle size distribution (e.g., Hansen and Travis, 1974) with the effective variance fixed to $\nu_{eff}$ = 0.3 (e.g., Smith and Wolff, 2014). The effective radius ($r_{eff}$) and the column optical depth of dust ($\tau$) were left as free parameters.

Regarding the shape of the particles, for Navcam Sun-pointing observation comparisons, spherical particles were considered for simplicity, as particle's shape has negligible effects in the forward scattering region of the phase function (Kaufman et al., 1994; Pollack et al., 1995). Conversely, for backscatter simulations, spheroidal particles were assumed for modelling the light scattering by Martian dust aerosol into this region (Mishchenko et al., 1997; Dluglach et al., 2002; Clancy et al., 2003; Dubovik et al., 2006; Merikallio et al., 2011; Nouisiainen et al., 2011). For calculating the scattering properties of the ensemble of spheroidal particles with MOPSMAP, we considered a mixture of prolate (elongated) and oblate (flat) spheroids following a two-parameter modified lognormal aspect ratio distribution (Kandler et al., 2007), being the aspect ratio defined as the ratio between the longest and shortest axis (Gasteiger and Wiegner, 2018). The distribution's width was fixed to $\sigma_{ar}$ = 0.60 as typical value for mineral-dust like particles (e.g., Kandler et al., 2007; 2011); and the parameter $\epsilon'_0$, which indicates the location of the maximum of the aspect ratio distribution, was left as free parameter.

### 3.2 Double Henyey-Greenstein phase functions
For comparisons with Navcam backscatter observations, we used three-term Double Henyey-Greenstein (DHG) analytical single scattering phase functions to evaluate the angular scattering behaviour of dust at intermediate and large angles during the storm. DHG parameters controlling the forward scattering ($g_1$), backward scattering ($g_2$), and the forward-backward ratio ($\alpha$) were varied to simulate different aerosol phase functions (Zhang and Li, 2016). The single scattering albedo was fixed to $\omega_0$ = 0.975, as an average value of results reported in Wolff et al. (2009) from surface-orbit observations and particularised for 650 nm. Due to computation time constrains, we limited the number of free parameters in the radiative transfer code in this DHG approach, and thus the column dust opacity was pre-defined in those simulations.

### 3.3 Retrieval procedure
A brute-force iterative retrieval scheme was implemented based in the comparison of the observed and modelled sky brightness as a function of the parameters characterising the airborne dust particles. Those parameter values generating the best fitting curve under a lowest mean quadratic deviation $\chi^2$ criterion were considered the outputs of the retrieval, and the uncertainties were evaluated around the best-fitting location in the free parameter space.

For Navcam Sun-pointing observations, the sky radiance was sampled along a diagonal direction from scattering angles $\theta = 4°$ to $30°$, with steps of $1°$, in order to skip the saturated pixels of the solar disc region and limit the contributions from instrumental stray and scattered light (Soderblom et al., 2008; Smith and Wolff, 2014; Chen-Chen et al., 2019a). The radiative transfer modelled curves were generated for dust aerosol properties loaded from MOPSMAP and covering a wide range of dust scenarios during the GDS: $r_{eff}$ was varied from 0.50 to 10.00 um with 0.02 steps, and τ was evaluated from 0.30 to 10.00 with a step of 0.02 (Figure 2).

When evaluating the backward scattering observations, both MOPSMAP and DHG aerosol modelling approaches were used (Figure 3). In these observations, the retrieval of sky brightness as a function of the scattering angle from $\theta = 90°$ to about $160°$, depending on the observation geometry, was performed along the solar principal plane.

For radiative transfer simulations using the MOPSMAP aerosol database, a mixture of spheroidal particles following the previously described aspect ratio distribution was assumed; with parameter $\epsilon'_0$ varying from 1.2 to 4.5 with a step of 0.1. In these simulations the effective radius and column dust opacity were also left as free parameters. For the purpose of limiting the computation time, the evaluation interval for $r_{eff}$ and τ were re-defined with respect to the MOPSMAP Sun-pointing case. First, the quantities ($r_{eff}$, τ) were interpolated for the particular sol from Navcam Sun-pointing results. Subsequently, the new evaluation intervals were defined as follows: the effective radius was varied from 0.5 to 2.0 times the interpolated $r_{eff}$; whereas the opacity was evaluated from 0.5 to 1.5 times the interpolated τ. In both cases, a step of 0.05 was selected.

Within the DHG approach for backscatter radiative transfer simulations, the set of parameters ($g_1$, $g_2$, α) defining the shape of the single scattering phase functions were left as free parameters. The $g_1$ parameter was iterated from 0.50 to 1.00 with 0.01 steps, $g_2$ was varied from $-g_1$ to $+g_1$ in 50 divisions to prevent the backscatter lobe from being greater than the forward one (Zhang and Li, 2016), and α was sampled from 0.50 to 1.00 with 0.01 steps. Regarding the column dust optical depth, the τ values previously derived from Sun-pointing observations were used as an input.

We summarise on Table 1 the aerosol models used in this work, their associated free parameters and their respective evaluation intervals.

The $\chi^2$ values were derived in a successive manner from comparisons between the observed sky brightness as a function of the scattering angle $(I/F)(\theta)_{obs}$, and the modelled sky brightness curves $(I/F)(\theta)_{model}$ calculated with combinations of the free parameters. A standard $\chi^2$ method was used in the form of

$$\chi^2 = \sum_{i=1}^{N} \left( \frac{(I/F)_{obs_i} - (I/F)_{model_i}}{\sigma_i \cdot (I/F)_{obs_i}} \right)^2 \qquad (1)$$

where for the $N$ sampled points along the curve the modelled and observed radiance factors were compared using a least squares quadratic error criterion. The variance $\sigma_i$ is associated to the absolute calibration uncertainty of MSL Navcams. Due to the severe atmospheric dust loading conditions, for this work this uncertainty was estimated by comparing Navcam and Mastcam images obtained under similar pointing and at similar time (see Soderblom et al. (2008) for a detailed description). We compared the radiance of the same sky-region in Gale Crater north rim pointing images retrieved by Navcam and Mastcam-L0R filter for sols ranging from 2025 (pre-GDS) to 2225 (post-GDS). The resulting average difference between the two imagers was 10%. The absolute radiance calibration uncertainty reported for Mastcam in Bell et al. (2017) was about 10%, therefore the Navcams' absolute radiance error of $\sigma_i = 20\%$ was considered for calculating the uncertainties of the retrieval procedure.

Finally, the set of parameters generating the minimum $\chi^2$ value were considered the solutions of the retrieval. The uncertainty associated to each parameter was calculated from the 68.3% confidence region (1-σ error).

3.4 Sensitivity study

The sensitivity of the retrieval procedure for Navcam Sun-pointing observations can be derived from the $\chi^2$ maps of the $\tau$-$r_{eff}$ parameter-space in Figure 2. In this figure, the shape of the best fitting region shows that the retrieval is more sensitive to the input dust optical depth (horizontal axis) than to the effective radius. This is mainly due to the effects that each parameter has on the modelled curves of the forward scattering peak: while the effective radius parameter controls the curvature of the forward scattering lobe, the input column dust optical depth sets the overall radiance factor levels of the simulated curve; the latter being particularly relevant in high dust opacity scenarios.

For model-observation comparisons in the backscattering case, due to computational time limitations, the number of free parameters were limited to three in both cases, ($r_{eff}$, $\tau$, $\epsilon'_0$) for MOPSMAP and ($g_1$, $g_2$, $\alpha$) for DHG models; therefore some assumptions were made on the rest of intervening parameters. In the next lines we evaluate the robustness of the results when these parameters are varied.

In the MOPSMAP model, the variation in the particle shape aspect-ratio parameter $\epsilon'_0$ was evaluated for fixed values of dust particle size parameter $r_{eff}$ corresponding to the outputs of Sun-pointing retrievals. An average absolute difference of 12% was obtained, with maximum deviations of about +50% and -40% for GDS sols 2125 and 2148, respectively. For post-storm sols, no differences were obtained in the retrieval comparisons. We have also analysed the variations in the retrieved parameters when, instead of the log-normal aspect ratio distribution, oblate spheroids with a single aspect ratio were considered. In such an extreme case, average differences of about +50% were retrieved for $r_{eff}$ for GDS sols, while no variations were obtained for post-storm sols. The column dust opacity showed an overall decrease of about 15%, and the best-fitting aspect ratio values decreased about 15% during the GDS, while incrementing around 11% in post-storm sols.

In DHG simulations, we have evaluated the sensitivity of DHG parameters to variations in the input optical depth (+15%, -15%), and to a change in the single scattering albedo $\omega_0$, from 0.975 to 0.930 (e.g., Pollack et al., 1995). A reduction in the input dust opacity resulted in a generalised decline in all DHG parameters: around 10% for $g_1$ and $\alpha$, and up to 45% for $g_2$. These behaviours were similar when dust opacity was incremented 15%: again, no significant variations were detected for $g_1$ and $\alpha$, with maximum differences below 7%, while for the backscattering parameter differences of more than 50% were retrieved. Finally, the reduction of the $\omega_0$ was translated into considerable variations in all parameters, especially relevant for high-opacity sols: $g_1$ and $\alpha$ differed about 20% from the nominal case, while a decline in $g_2$ results from 30% (post-storm sols) to about 100% (dusty sols) was detected

**Table 1. Aerosol model parameters for radiative transfer simulations**

| Observation | Aerosol model | Single scat. albedo, $\omega_0$ | Phase function, P(θ) Parameters | Phase function, P(θ) Range |
|---|---|---|---|---|
| Sun-pointing | MOPSMAP | Calculated | Size distribution effective radius ($r_{eff}$), dust column opacity ($\tau$) | $r_{eff}$: 0.50 to 10.00 µm, step of 0.02<br>$\tau$: 0.30 to 10.00, step of 0.02 |
| Backward scattering region | MOPSMAP | Calculated | Aspect ratio distribution parameter ($\epsilon'_0$), size distribution effective radius ($r_{eff}$), dust column opacity ($\tau$) | $\epsilon'_0$: 1.2 to 4.5, step of 0.1<br>$r_{eff}$: 0.5 to 2.0 times the $r_{eff}$ derived in Sun-pointing observations, step of 0.05<br>$\tau$: 0.5 to 1.5 times the $\tau$ derived in Sun-pointing observations, step of 0.05 |
| Backward scattering region | Double Henyey-Greenstein | Fixed to: 0.975 | Forward scattering ($g_1$), backward scattering ($g_2$), and ratio ($\alpha$) | $g_1$: 0.50 to 1.00, step of 0.01<br>$g_2$: -$g_1$ to +$g_1$, 50 divisions<br>$\alpha$: 0.50 to 1.00, step of 0.01 |

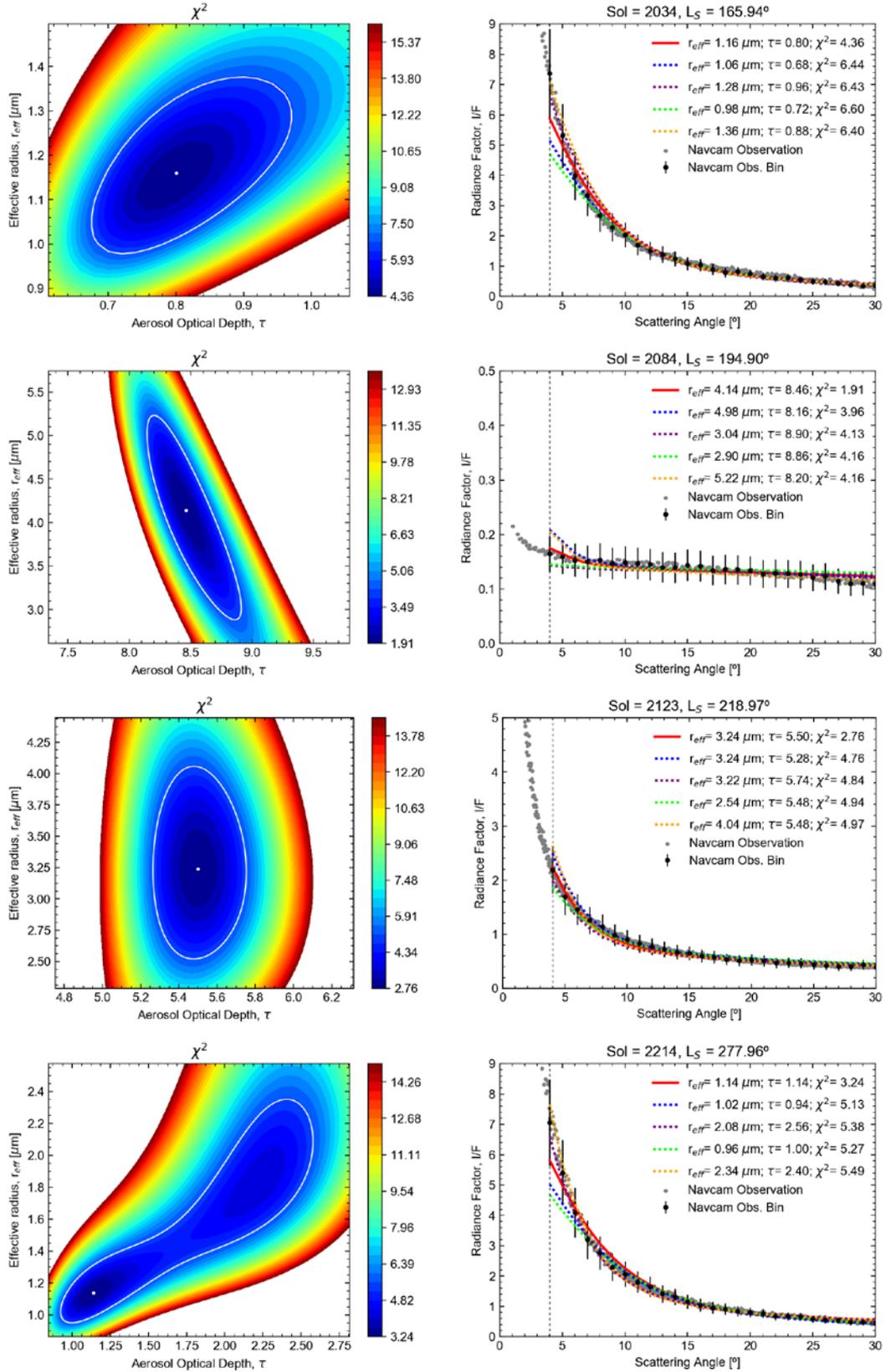

**Fig. 2. Sun-pointing observations and model comparisons.** *Results for sols 2034, 2084, 2123, and 2214 covering different opacity scenarios. Left: $\chi^2$ maps from model-observation comparisons in the $\tau$-$r_{eff}$ parameter space. The point of minimum $\chi^2$ and the contour of the 68.3% confidence interval limit are indicated. Right: Navcam observed sky radiance (gray) and binned data (black) with the calibration uncertainty (20%). The best fitting curve (red) and a family of good fits formed of extremal points on the contour line when moving horizontally and vertically with respect to minimum $\chi^2$ point, are also provided.*

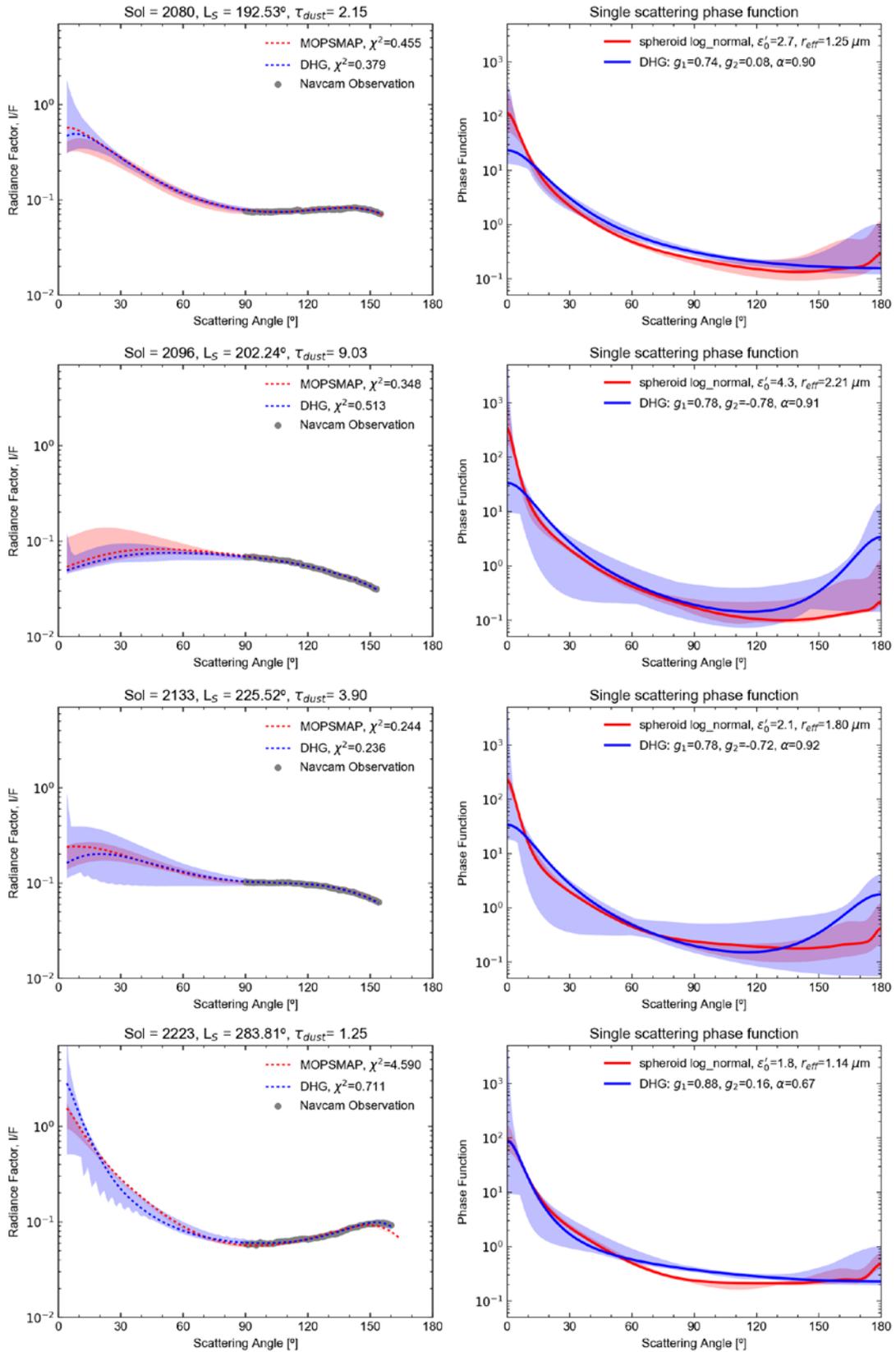

***Fig. 3. Backward scattering region observations and model comparison.*** *Results for sols 2080, 2096, 2133 and 2223 covering different dust loading scenarios during and after the 2018/MY34 GDS are shown in this panel. Left: the best fitting sky brightness curves simulated with MOPSMAP (red) and DHG (blue) aerosol models are plotted together with the Navcam observation (gray). Right: the single scattering phase function generating the corresponding best fitting curves. The shaded areas represent all the corresponding solutions contained within the 68.3% confidence interval of the $\chi^2$ retrieval.*

## 4. Results and discussion

The methodology described in the previous Section was followed to retrieve the aerosol model parameters generating the best-fitting simulations to Navcams' sky radiance observations. In this Section 4, the retrieval outputs are presented and a discussion is provided regarding the behaviour of dust aerosol properties during the 2018/MY34 GDS. The results are provided on Table A2 in the Appendix A of this manuscript.

4.1 Dust column opacity

The retrieved column opacities are shown in Figure 4. The first observation is sol 2034 ($L_S = 165º$), about 40 sols before the arrival of the storm to Gale Crater; the corresponding atmospheric opacity of 0.80 is within the usual range for that period (Lemmon et al., 2015). The following 5 observations cover the arrival of the dust storm to the rover site, during which a steep increase of column dust opacity is derived; with values escalating from 2.15 in sol 2080 to $\tau > 9$ around sol 2089, the maximum opacity values detected in this study. Afterwards, a gradual decay in dust opacity can be appreciated, starting around sol 2097 ($L_S = 203º$) with $\tau = 8.00$ to optical depth levels of ~ 1.7 in sol 2165 ($L_S = 246º$). Finally, the dust loading scenario recovers pre-storm levels with $\tau$ ~ 1.30 to 1.15 in the last sols of this data-set ($L_S$ ~ 280º).

The comparison to MSL Mastcam direct Sun-imaging opacity measurements (Lemmon et al., 2019) shows that, during GDS sols, Navcam retrievals are generally higher: an average difference of about 35% was obtained, with a maximum of 50% for sol 2118, whereas the pre- and post-storm differences reduce down to about 8% and 21%, respectively. Possible rationales to this deviation may be found, first, on the time at which observations were retrieved. Navcam Sun-pointing observations were performed within the 11-12h and 14-15h LTST frames, while Mastcam measurements were mainly obtained in the afternoon (14-17h LTST) (Lemmon et al., 2019). The analysis of the diurnal opacity variations during the 2018/MY34 GDS reported by previous authors showed larger morning opacities than afternoon ones (Guzewich et al., 2019; Smith et al., 2019). A similar behaviour was reported at Viking lander sites during the 1977/MY12 event by Colburn et al. (1989), who related the morning-afternoon opacity differences to the condensation of water upon suspended dust particles during the colder portions of the sol. Another possible source of differences in may be due to the presence of dust on Navcams' lenses, as suggested by the existing deviations with respect to Mastcam for sols during and outside the GDS. This thin layer of dust contributes with additional opacity, e.g. a "$\tau_{window}$" (Lemmon et al., 2015), which shall be characterised and removed from the retrievals. When comparing the retrieved optical depth pattern with respect to Navcam line-of-sight extinction results (Smith et al., 2019), the dust column opacities reported in this work do not show the double-peaked structure reported in that study. However, Smith et al. (2019) associated their retrieved pattern to variations and effects constrained to the near surface which would present negligible effects in the column opacity.

4.2 Particle size

The dust particle sizes derived with Sun-pointing images are also provided in Figure 4. During the 2018/MY34 GDS, the particle size distribution effective radius shows an overall variation from values of about 1.15 μm, corresponding to pre- and post-storm sols, up to a maximum of $r_{eff}$ = 4.14 μm, inferred for sol 2084 (close to GDS peak of opacity). These results show that during severe dust storms larger particles are lifted. They provide further evidence of the positive correlation between dust opacity and particle size, extend the range of applicability, and present an overall agreement with previous studies (Kahre et al., 2008; Vincendon et al., 2009; Smith and Wolff, 2014; Vicente-Retortillo et al., 2017; McConnochie et al., 2018; Chen-Chen et al., 2019a). It is worth pointing out that the variation in dust opacity observed within part of the decay phase, spanning from sol 2089 to 2132 (last Navcam Sun-pointing observation during the storm), is not followed by the particles' size, as this shows a steady behaviour around $r_{eff}$ ~ 3.5 μm in this period.

The comparison of Navcam retrievals with Mastcam near-Sun results (Lemmon et al., 2019) returns an average difference of 17%. However, large-sized particles with $r_{eff}$ values of > 5.0 and up to almost 9.0 μm such as those reported in Lemmon et al. (2019) were not retrieved in our study, with a particle size population ranging from 1.14 to 4.14 μm. In addition, the second peak around sol 2010-2120 retrieved by Mastcam, with particle sizes ranging 3-5 μm and reaching up to 6 μm, was not detected by Navcams, which reported sizes of ~3.5 μm. Possible rationales to the deviations may be found in the time of the retrieval, as abovementioned for the opacity case. Particularly, during the second peak period, Navcam observations were obtained in the 11-12h LTST frame, whereas Mastcam retrievals with sizes > 4 μm were made between 15 and 16h LTST.

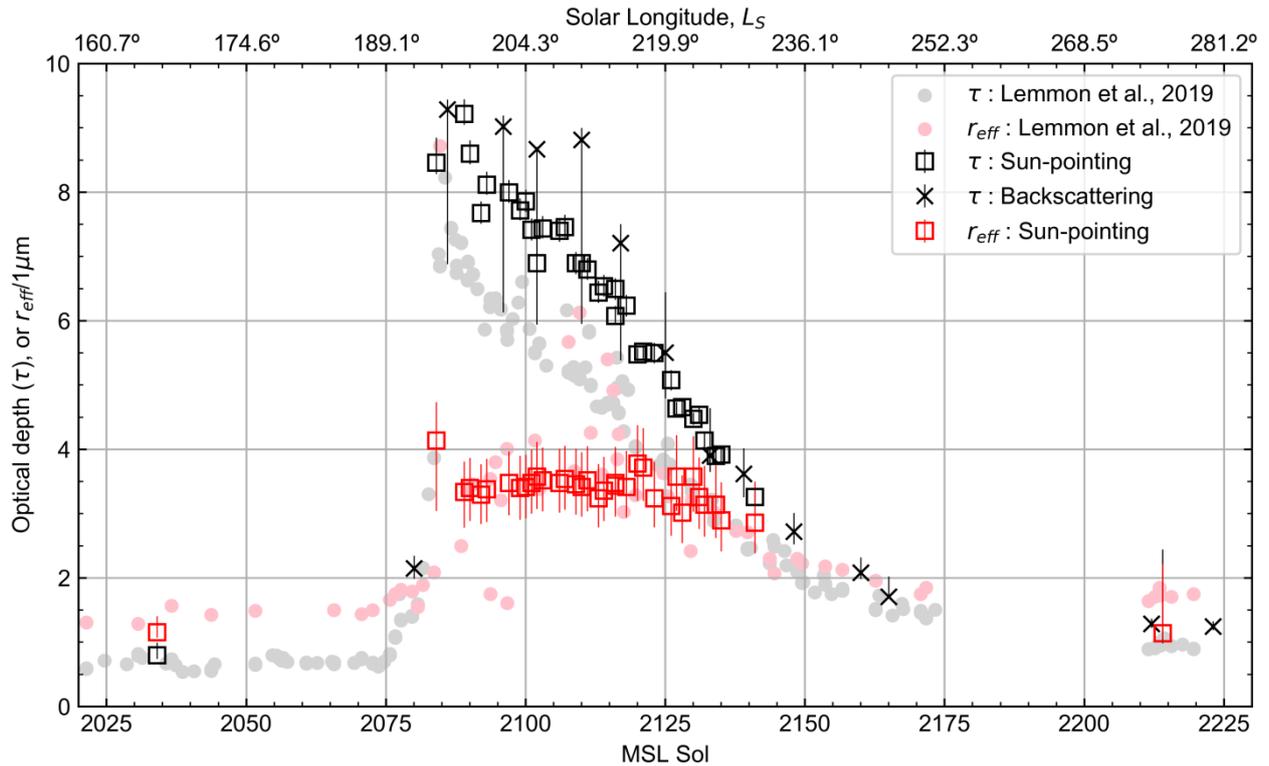

*Fig. 4. Dust optical depth and particle size during 2018/MY34 GDS. Values for dust column opacity (black) and particle size distribution effective radius (red) retrieved in this study using Sun-pointing (square) and backward-scattering (cross) Navcam observations during the 2018/MY34 global dust storm. The existing data gap covering sols 2173 to 2211 ($L_S$ ~252º to 276º) is due to a rover anomaly (Lemmon et al., 2019). For comparison purposes, MSL Mastcam retrievals reported in Lemmon et al. (2019) are provided.*

4.3 Particle shape
The behaviour of the shape of dust particles during the GDS is illustrated on Figure 5 (left panel). In this figure, we show the results for the $\epsilon_0'$ parameter of the mixture of prolate and oblate spheroids with a modified log-normal aspect ratio distribution using MOPSMAP database. The variation of particles' non-sphericity parameter with dust optical depth presents an average value of around 2.9 during the GDS, with maximum values reaching $\epsilon_0' > 4.0$ at peak atmospheric dust opacity sols, whereas for post-storm conditions the value drops down to 1.8. The resulting modified lognormal aspect ratio distributions (Kandler et al., 2007; Gasteiger and Wiegner, 2018) are plotted on the right panel of Figure 5 for those situations, in addition to the minimum (1.6) and maximum (4.4) values obtained in the retrieval.

Results of the particles' shape analysis are, on one hand, in line with main findings from previous studies of particle scattering properties; pointing that aspect ratio distributions of spheroidal particles in which high aspect ratio (> 2) spheroids are predominant can be good approaches for modelling the light scattering by mineral dust aerosol particles (Clancy et al., 2003; Dubovik et al., 2006; Nouisiainen et al., 2006, 2011; Merikallio et al., 2011; 2013). On the other hand, retrievals are also in good agreement with conclusions drawn from Viking landers 1977/MY12 GDS observations: analysis of sky brightness imaging data suggested non-spherical plate-like particles (Pollack et al., 1979); and the numerical modelling of the decay phase of Martian dust storms reported that particles should be disc-shaped in order to match the opacity decay rates observed by the Viking landers (Murphy et al., 1990).

4.4 Single scattering phase function
The retrieved DHG parameters ($g_1$, $g_2$, α) using Navcam backscatter observations during the 2018/MY34 GDS are provided in Figure 6. These are compared with retrievals reported in Chen-Chen et al. (2019b) covering the same solar longitudes ($L_S$ = 175º to 280º) for previous Martian years. The larger uncertainty bars featured in this study are due to the non-availability of sky brightness observations covering simultaneously the forward and backward scattering angle regions, i.e., "sky surveys".

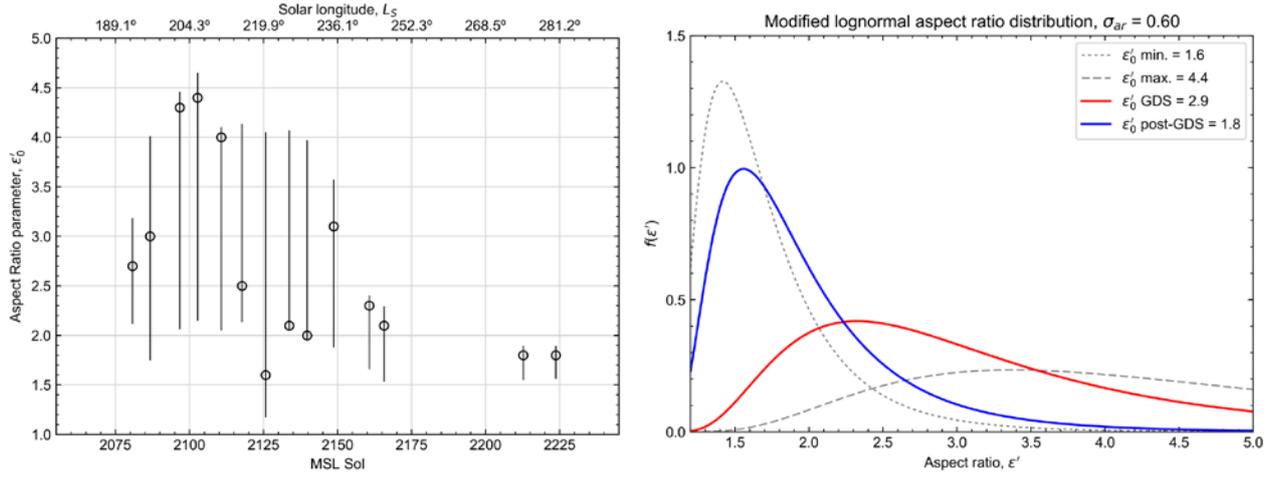

**Fig. 5. Results for particle shape.** *Left: Evolution of the aspect ratio parameter $\epsilon'_0$ of the aspect ratio distribution during the 2018/MY34 GDS. Right: Aspect ratio distributions following a modified lognormal distribution for a mixture of prolate and oblate spheroids (Kandler et al., 2007; Gasteiger and Wiegner, 2018), variance parameter fixed to $\sigma_{ar} = 0.60$, for 4 different scenarios: minimum and maximum derived aspect ratio parameter values, and the mean value parameter values during and after the GDS.*

The best-fitting values of the parameter $g_1$ derived from Navcam observations at high-opacity sols during the GDS return a mean value of $g_1 = 0.79 \pm 0.07$; while for post-storm observations on sols 2212 and 2223, this parameter increases to $g_1 = 0.89$. On Figure 6, it can be appreciated that seasonal values of $g_1$ derived for previous Martian Years are mostly located within the 0.9 to 1.0 range, being the average for that specific solar longitude period of $g_1 = 0.93 \pm 0.06$. Regarding the parameter describing the backward scattering behaviour, during high opacity sols the retrieved mean value is $g_2 = -0.39 \pm 0.44$; which contrasts with the average of 0.18 obtained for post-storm sols, and as well as with retrievals for MY 31 to 33 for the same solar longitude period, with average $g_2$ of $0.20 \pm 0.14$ (Chen-Chen et al., 2019b). Results obtained for the forward-backward scattering ratio (α) also show noticeable differences, with average value of $\alpha = 0.82 \pm 0.14$ for observations retrieved during high-opacity sols, and values of around 0.64 for post-storm sols; being the latest ones more similar to the those retrieved in years with no dust storms, where the average of $\alpha = 0.68 \pm 0.10$.

Based on these results, the asymmetry parameter (g) can be calculated from the derived DHG parameters (Zhang and Li, 2016). During the GDS, an average value of $g = 0.601 \pm 0.108$ is obtained, while values for the post-storm sols are of $g = 0.642$; which is more similar to the overall average of $g = 0.710 \pm 0.065$ reported in Chen-Chen et al. (2019b) for previous Martian Years and from previous studies (e.g., Wolff et al., 2009). Finally, the interrelationships between the DHG parameters derived during the GDS are in good agreement with trends previously reported in Chen-Chen et al. (2019b); showing positive correlation for $g_1$-$g_2$ parameters, and negative correlations for $g_1$-α, and $g_2$-α, and populating the extreme regions of the evaluated range.

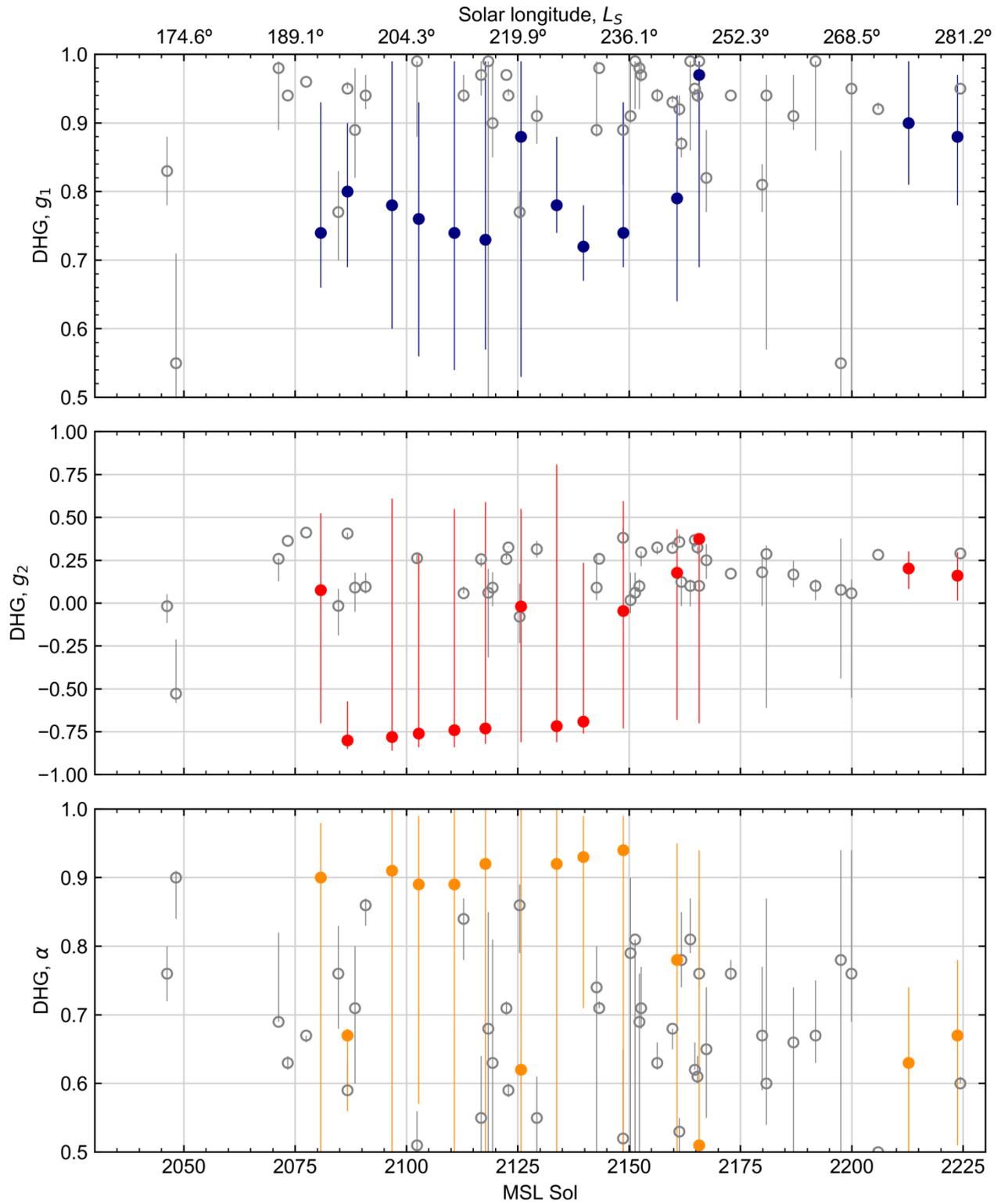

*Fig. 6. Double Henyey-Greenstein phase function parameters during 2018/MY34 GDS.* Results of DHG analytical single scattering phase function parameters $g_1$ (blue), $g_2$ (red) and $\alpha$ (yellow) using Navcam backscatter observations during the 2018/MY34 GDS are provided in these graphs. For comparison purposes, the values reported in Chen-Chen et al. (2019b) for previous MY are also provided (circles).

# 5. Conclusion

We have derived the atmospheric dust opacity and characterised the aerosol particle properties during the 2018/MY34 Martian global dust storm using MSL *Curiosity* rover Navigation Cameras observations. A total of 49 Sun-pointing and backward scattering region observation sequences, spanning from sol 2034 to 2223 ($L_S$ ~ 166º to 284º), have been evaluated covering the onset, outburst and decay of the dust storm event.

The angular decay of sky brightness observed by the Navcams was compared with radiative transfer simulations in order to retrieve the aerosol parameters that generate the best-fitting curve. The MOPSMAP aerosol database (Gasteiger and Wiegner, 2018) with pre-computed single scattering properties for different particles have been used for modelling the Martian dust aerosol, together with the three term Double Henyey-Greenstein expression for analytical single scattering phase functions.

MSL Navcam retrievals for column opacity during the 2018/MY34 global dust storm show a steep increase, within 5 sols, from usual seasonal dust opacity levels of 1.2 up to peaks of τ > 9, followed by a decay phase down to 1.7 in a period of approximately 80 sols, and concluding with post-storm observations of τ ~ 1.2, back to common optical depth levels for that period ($L_S$ ~ 280º) reported for previous Martian Years (Lemmon et al., 2015). Dust particle sizes from 1.2 to about 4.1 μm have been derived with Sun-pointing observations, showing a positive correlation between dust opacity and particle sizes, and supporting the evidence of lifting of larger dust particles during severe dust storms suggested in previous studies (e.g., Kahre et al., 2008; Vincendon et al., 2009; Smith and Wolff, 2014; Vicente-Retortillo et al., 2017). Comparisons with MSL Mastcam opacity measurements and particle size retrievals reported in Lemmon et al. (2019) for the GDS show average differences of 34% and 17% respectively, which can be used to constrain the overall uncertainties in the retrievals.

Navcam backward scattering region observations have been used to evaluate the behaviour of the single scattering phase function and to characterise the shape of the particles. Double Henyey-Greenstein parameters show significant differences when GDS retrieval values ($g_1$ = 0.79±0.07, $g_2$ = -0.39±0.14, and α = 0.82±0.14) are compared to post-storm sols or results for previous Martian Years ($g_1$ = 0.93±0.06, $g_2$ = 0.20±0.14, α = 0.68±0.10; Chen-Chen et al., 2019b). These outcomes result in phase functions featured with asymmetry parameter values of g = 0.601±0.108 for high atmospheric dust loading scenarios lower than for non-dusty periods, with g = 0.710±0.065. Regarding the shape of the particles, we have analysed the performance of a modified log-normal aspect ratio distribution for a mixture of spheroids. The best-fitting parameters in model-observation comparisons suggest a trend for more elongated particles during high atmospheric dust loading sols, with aspect ratios = 2.8±0.9 for high-opacity sols, in contrast to values of 1.8 during post-storm sols. The systematic and continuous characterisation of the behaviour of dust single scattering phase function and particle's shape performed in this work may provide further information in the interrelationship between particle properties and dust cycle dynamical processes, including its transport and deposition rates.

Future research prospects include the numerical modelling of the onset and development local and regional dust storms. In combination with MSL detailed atmospheric measurements (Viúdez-Moreiras et al., 2019), outcomes of this study merits to be used in mesoscale model simulations for evaluating the outburst and behaviour of local and regional-scale dust storms, and the interpretation of their effects in the atmospheric variables (Rafkin et al., 2016; Rafkin and Michaels, 2019).

**Acknowledgements**
This work was supported by the Spanish project AYA2015-65041-P (MINECO/FEDER, UE), Grupos Gobierno Vasco IT1366-19, and Diputación Foral de Bizkaia - Aula EspaZio Gela.

**Appendix A**

## Table A1. MSL Navcam 2018/MY34 GDS Observations

| MSL SOL | $L_S$ [deg] | Type | LTST | Sun Azim. [deg] | Sun Elev. [deg] | Files |
|---|---|---|---|---|---|---|
| 2034 | 165.94 | SUN | 14.97 | 282.96 | 44.295 | NRB_578066685EDR_F0692976SAPP07612M1.IMG |
| 2080 | 192.53 | BKWD | 16.94 | 265.82 | 16.22 | NLB_582157410EDR_D0701752NCAM00565M1.IMG, NLB_582157434EDR_D0701752NCAM00565M1.IMG, NLB_582157466 EDR_D0701752NCAM00565M1.IMG, NLB_582157490 EDR_D0701752NCAM00565M1.IMG, NRB_582157410EDR_D0701752NCAM00565M1.IMG, NRB_582157434EDR_D0701752NCAM00565M1.IMG, NRB_582157466 EDR_D0701752NCAM00565M1.IMG, NRB_582157490 EDR_D0701752NCAM00565M1.IMG |
| 2084 | 194.90 | SUN | 15.30 | 266.34 | 40.94 | NRB_582506466EDR_F0701752SAPP07612M1.IMG |
| 2086 | 196.15 | BKWD | 16.94 | 264.26 | 16.33 | NLB_582690066EDR_D0710060NCAM00565M1.IMG, NLB_582690091 EDR_D0710060NCAM00565M1.IMG, NLB_582690126 EDR_D0710060NCAM00565M1.IMG, NLB_582690153 EDR_D0710060NCAM00565M1.IMG, NRB_582690066EDR_D0710060NCAM00565M1.IMG, NRB_582690091 EDR_D0710060NCAM00565M1.IMG, NRB_582690126 EDR_D0710060NCAM00565M1.IMG, NRB_582690153 EDR_D0710060NCAM00565M1.IMG |
| 2089 | 197.82 | SUN | 11.19 | 102.38 | 77.53 | NRB_582935154EDR_F0710066NCAM00601M1.IMG, NRB_582935160EDR_F0710066NCAM00601M1.IMG |
| 2090 | 198.44 | SUN | 11.62 | 116.14 | 83.66 | NRB_583025551EDR_F0710228NCAM00600M1.IMG, NRB_583025564EDR_F0710228NCAM00600M1.IMG |
| 2092 | 199.67 | SUN | 11.70 | 126.46 | 84.40 | NRB_583203403EDR_F0710228NCAM00600M1.IMG, NRB_583203422EDR_F0710228NCAM00600M1.IMG |
| 2093 | 200.29 | SUN | 12.18 | 216.19 | 85.59 | NRB_583293980EDR_F0710570NCAM00600M1.IMG, NRB_583293980EDR_F0710570NCAM00600M1.IMG |
| 2096 | 202.24 | BKWD | 16.60 | 261.85 | 21.47 | NLB_583576614EDR_F0711330NCAM00565M1.IMG, NLB_583576639EDR_F0711330NCAM00565M1.IMG, NLB_583576671EDR_F0711330NCAM00565M1.IMG, NLB_583576696EDR_F0711330NCAM00565M1.IMG, NRB_583576614EDR_F0711330NCAM00565M1.IMG, NRB_583576639EDR_F0711330NCAM00565M1.IMG, NRB_583576671EDR_F0711330NCAM00565M1.IMG, NRB_583576696EDR_F0711330NCAM00565M1.IMG |
| 2097 | 202.76 | SUN | 12.85 | 249.00 | 76.73 | NRB_583651541EDR_F0711330NCAM00600M1.IMG, NRB_583651559EDR_F0711330NCAM00600M1.IMG |
| 2099 | 203.98 | SUN | 12.34 | 224.76 | 82.94 | NRB_583827239EDR_F0711586NCAM00600M1.IMG, NRB_583827258EDR_F0711586NCAM00600M1.IMG |
| 2100 | 204.61 | SUN | 12.88 | 247.15 | 76.00 | NRB_583918016EDR_F0711586NCAM00600M1.IMG, NRB_583918034EDR_F0711586NCAM00600M1.IMG |
| 2101 | 205.21 | SUN | 12.02 | 183.19 | 84.49 | NRB_584003627EDR_F0711586NCAM00600M1.IMG, NRB_584003646EDR_F0711586NCAM00600M1.IMG |
| 2102 | 205.87 | SUN | 13.75 | 255.62 | 63.29 | NRB_584098807EDR_F0711818NCAM00600M1.IMG, NRB_584098825EDR_F0711818NCAM00600M1.IMG |
| 2102 | 205.96 | BKWD | 16.99 | 260.13 | 15.71 | NLB_584110769EDR_D0711818NCAM00565M1.IMG, NLB_584110795EDR_D0711818NCAM00565M1.IMG, NLB_584110828EDR_D0711818NCAM00565M1.IMG, NLB_584110854EDR_D0711818NCAM00565M1.IMG, NRB_584110769EDR_D0711818NCAM00565M1.IMG, NRB_584110795EDR_D0711818NCAM00565M1.IMG, NRB_584110828EDR_D0711818NCAM00565M1.IMG, NRB_584110854EDR_D0711818NCAM00565M1.IMG |
| 2103 | 206.43 | SUN | 11.52 | 130.81 | 80.64 | NRB_584179313EDR_F0711818NCAM00600M1.IMG, NRB_584179338EDR_F0711818NCAM00600M1.IMG |
| 2106 | 208.31 | SUN | 11.89 | 166.10 | 83.04 | NRB_584447048EDR_F0712350NCAM00600M1.IMG, NRB_584447072EDR_F0712350NCAM00600M1.IMG |
| 2107 | 208.93 | SUN | 11.95 | 173.74 | 82.94 | NRB_584536053EDR_F0712350NCAM00600M1.IMG, NRB_584536078EDR_F0712350NCAM00600M1.IMG |
| 2109 | 210.18 | SUN | 11.68 | 148.64 | 81.16 | NRB_584712661EDR_F0712876NCAM00600M1.IMG, NRB_584712679EDR_F0712876NCAM00600M1.IMG |
| 2110 | 210.81 | SUN | 12.02 | 182.02 | 82.17 | NRB_584802682EDR_F0712876NCAM00600M1.IMG, NRB_584802700EDR_F0712876NCAM00600M1.IMG |
| 2110 | 210.93 | BKWD | 16.69 | 258.23 | 20.13 | NLB_584819957EDR_D0712876NCAM00565M1.IMG, NLB_584819983EDR_D0712876NCAM00565M1.IMG, NLB_584820016EDR_D0712876NCAM00565M1.IMG, NLB_584820041EDR_D0712876NCAM00565M1.IMG, NRB_584819957EDR_D0712876NCAM00565M1.IMG, NRB_584819983EDR_D0712876NCAM00565M1.IMG, NRB_584820016EDR_D0712876NCAM00565M1.IMG, |

| MSL SOL | L$_S$ [deg] | Type | LTST | Sun Azim. [deg] | Sun Elev. [deg] | Files |
|---|---|---|---|---|---|---|
| | | | | | | NRB_584820041EDR_D0712876NCAM00565M1.IMG |
| 2111 | 211.44 | SUN | 12.01 | 181.32 | 81.95 | NRB_584891443EDR_F0712876NCAM00600M1.IMG, NRB_584891461EDR_F0712876NCAM00600M1.IMG |
| 2113 | 212.69 | SUN | 12.09 | 189.14 | 81.49 | NRB_585069308EDR_F0712876NCAM00600M1.IMG, NRB_585069338EDR_F0712876NCAM00600M1.IMG |
| 2114 | 213.33 | SUN | 12.26 | 203.96 | 80.46 | NRB_585158750EDR_F0712876NCAM00600M1.IMG, NRB_585158774EDR_F0712876NCAM00600M1.IMG |
| 2116 | 214.56, 214.63 | SUN SAPP | 11.30, 13.92 | 132.07, 250.33 | 76.14, 60.26 | NRB_585332764EDR_F0712956NCAM00600M1.IMG, NRB_585332788EDR_F0712956NCAM00600M1.IMG, NRB_585342484EDR_F0713370SAPP07612M1.IMG |
| 2117 | 215.34 | BKWD | 16.94 | 256.49 | 16.59 | NLB_585442389EDR_D0720000NCAM00565M1.IMG, NLB_585442414EDR_D0720000NCAM00565M1.IMG, NLB_585442447EDR_D0720000NCAM00565M1.IMG, NLB_585442472EDR_D0720000NCAM00565M1.IMG, NRB_585442389EDR_D0720000NCAM00565M1.IMG, NRB_585442414EDR_D0720000NCAM00565M1.IMG, NRB_585442447EDR_D0720000NCAM00565M1.IMG, NRB_585442472EDR_D0720000NCAM00565M1.IMG |
| 2118 | 215.85 | SUN | 12.14 | 190.70 | 80.23 | NRB_585513461EDR_F0720000NCAM00600M1.IMG, NRB_585513461EDR_F0720000NCAM00600M1.IMG |
| 2120 | 217.20 | SUN | 15.50 | 254.66 | 37.58 | NRB_585703459EDR_F0720386NCAM00600M1.IMG, NRB_585703471EDR_F0720386NCAM00600M1.IMG |
| 2121 | 217.82 | SUN | 15.11 | 253.44 | 43.05 | NRB_585790821EDR_F0720386NCAM00600M1.IMG, NRB_585790833EDR_F0720386NCAM00600M1.IMG |
| 2123 | 218.97 | SUN | 10.58 | 118.45 | 66.54 | NRB_585951630EDR_F0720386NCAM00600M1.IMG, NRB_585951666EDR_F0720386NCAM00600M1.IMG |
| 2125 | 220.41 | BKWD | 16.88 | 254.65 | 17.37 | NLB_586152541EDR_D0720386NCAM00565M1.IMG, NLB_586152566EDR_D0720386NCAM00565M1.IMG, NLB_586152598EDR_D0720386NCAM00565M1.IMG, NLB_586152623EDR_D0720386NCAM00565M1.IMG, NRB_586152541EDR_D0720386NCAM00565M1.IMG, NRB_586152566EDR_D0720386NCAM00565M1.IMG, NRB_586152598EDR_D0720386NCAM00565M1.IMG, NRB_586152623EDR_D0720386NCAM00565M1.IMG |
| 2126 | 220.96 | SUN | 13.54 | 241.36 | 64.66 | NRB_586228992EDR_F0720920NCAM00600M1.IMG, NRB_586229016EDR_F0720920NCAM00600M1.IMG |
| 2127 | 221.65 | SUN | 15.50 | 252.84 | 37.47 | NRB_586325018EDR_F0720920NCAM00600M1.IMG, NRB_586325036EDR_F0720920NCAM00600M1.IMG |
| 2128 | 222.26 | SUN | 14.30 | 248.10 | 54.27 | NRB_586409385EDR_F0721286NCAM00600M1.IMG, NRB_586409409EDR_F0721286NCAM00600M1.IMG |
| 2130 | 223.57 | SUN | 15.57 | 252.60 | 36.28 | NRB_586591679EDR_F0721286NCAM00600M1.IMG, NRB_586591696EDR_F0721286NCAM00600M1.IMG |
| 2131 | 224.18 | SUN | 14.43 | 247.80 | 52.29 | NRB_586676260EDR_F0721286NCAM00600M1.IMG, NRB_586676284EDR_F0721286NCAM00600M1.IMG |
| 2132 | 224.84 | SUN | 15.13 | 250.87 | 42.47 | NRB_586767636EDR_F0721316NCAM00600M1.IMG, NRB_586767666EDR_F0721316NCAM00600M1.IMG |
| 2133 | 225.52 | BKWD | 16.60 | 252.82 | 21.35 | NLB_586861885EDR_D0721316NCAM00565M1.IMG, NLB_586861910EDR_D0721316NCAM00565M1.IMG, NLB_586861941EDR_D0721316NCAM00565M1.IMG, NLB_586861966EDR_D0721316NCAM00565M1.IMG, NRB_586861885EDR_D0721316NCAM00565M1.IMG, NRB_586861910EDR_D0721316NCAM00565M1.IMG, NRB_586861941EDR_D0721316NCAM00565M1.IMG, NRB_586861966EDR_D0721316NCAM00565M1.IMG |
| 2134 | 226.13 | SUN | 15.37 | 251.03 | 38.99 | NRB_586946118EDR_F0721316NCAM00600M1.IMG, NRB_586946143EDR_F0721316NCAM00600M1.IMG |
| 2135 | 226.75 | SUN | 14.81 | 248.74 | 46.76 | NRB_587032858EDR_F0721316NCAM00600M1.IMG, NRB_587032858EDR_F0721316NCAM00600M1.IMG |
| 2139 | 229.39 | BKWD | 17.21 | 251.70 | 12.67 | NLB_587396940EDR_D0721316NCAM00565M1.IMG, NLB_587396965EDR_D0721316NCAM00565M1.IMG, NLB_587396998EDR_D0721316NCAM00565M1.IMG, NLB_587397023EDR_D0721316NCAM00565M1.IMG, NRB_587396940EDR_D0721316NCAM00565M1.IMG, NRB_587396965EDR_D0721316NCAM00565M1.IMG, NRB_587396998EDR_D0721316NCAM00565M1.IMG, NRB_587397023EDR_D0721316NCAM00565M1.IMG |
| 2141 | 230.63 | SUN | 15.32 | 249.21 | 39.47 | NRB_587567544EDR_F0721316NCAM00600M1.IMG, NRB_587567544EDR_F0721316NCAM00600M1.IMG |
| 2148 | 235.19 | BKWD | 16.44 | 249.69 | 23.55 | NLB_588193320EDR_D0721316NCAM00565M1.IMG, NLB_588193344EDR_D0721316NCAM00565M1.IMG, NLB_588193368EDR_D0721316NCAM00565M1.IMG, NLB_588193392EDR_D0721316NCAM00565M1.IMG, NRB_588193320EDR_D0721316NCAM00565M1.IMG, NRB_588193344EDR_D0721316NCAM00565M1.IMG, |

| MSL SOL | L$_S$ [deg] | Type | LTST | Sun Azim. [deg] | Sun Elev. [deg] | Files |
|---|---|---|---|---|---|---|
| | | | | | | NRB_588193368EDR_D0721316NCAM00565M1.IMG, NRB_588193392EDR_D0721316NCAM00565M1.IMG |
| 2160 | 243.00 | BKWD | 16.90 | 248.09 | 17.08 | NLB_589260715EDR_D0721980NCAM00565M1.IMG, NLB_589260739EDR_D0721980NCAM00565M1.IMG, NLB_589260772EDR_D0721980NCAM00565M1.IMG, NLB_589260796EDR_D0721980NCAM00565M1.IMG, NRB_589260715EDR_D0721980NCAM00565M1.IMG, NRB_589260739EDR_D0721980NCAM00565M1.IMG, NRB_589260772EDR_D0721980NCAM00565M1.IMG, NRB_589260796EDR_D0721980NCAM00565M1.IMG |
| 2165 | 246.24 | BKWD | 16.39 | 246.93 | 24.03 | NLB_589702904EDR_D0722410NCAM00565M1.IMG, NLB_589702929EDR_D0722410NCAM00565M1.IMG, NLB_589702961EDR_D0722410NCAM00565M1.IMG, NLB_589702986EDR_D0722410NCAM00565M1.IMG, NRB_589702904EDR_D0722410NCAM00565M1.IMG, NRB_589702929EDR_D0722410NCAM00565M1.IMG, NRB_589702961EDR_D0722410NCAM00565M1.IMG, NRB_589702986EDR_D0722410NCAM00565M1.IMG |
| 2212 | 276.78 | BKWD | 17.03 | 245.33 | 15.13 | NLA_593879667EDR_D0722464NCAM00565M1.IMG, NLA_593879691EDR_D0722464NCAM00565M1.IMG, NLA_593879724EDR_D0722464NCAM00565M1.IMG, NLA_593879748EDR_D0722464NCAM00565M1.IMG, NRA_593879667EDR_D0722464NCAM00565M1.IMG, NRA_593879691EDR_D0722464NCAM00565M1.IMG, NRA_593879724EDR_D0722464NCAM00565M1.IMG, NRA_593879748EDR_D0722464NCAM00565M1.IMG |
| 2214 | 277.96 | SAPP | 13.27 | 220.07 | 62.84 | NLA_594043397EDR_F0722464SAPP07612M1.IMG |
| 2223 | 283.81 | BKWD | 17.08 | 245.96 | 14.51 | NLA_594856798EDR_D0730055NCAM00565M1.IMG, NLA_594856823EDR_D0730055NCAM00565M1.IMG, NLA_594856856EDR_D0730055NCAM00565M1.IMG, NLA_594856880EDR_D0730055NCAM00565M1.IMG, NRA_594856798EDR_D0730055NCAM00565M1.IMG, NRA_594856823EDR_D0730055NCAM00565M1.IMG, NRA_594856856EDR_D0730055NCAM00565M1.IMG, NRA_594856880EDR_D0730055NCAM00565M1.IMG |

Observation sequences:
**BKWD:** sequence NCAM00565
**SAPP:** Surface Attitude Positioning and Pointing
**SUN:** sequence NCAM00600

Note: MSL mission sol number, solar longitude (in degrees), type of observation, local true solar time (in decimal format), solar azimuth and elevation angles (in degrees) with respect to North and local horizon. Files retrieved from PDS imaging node: *https://pds-imaging.jpl.nasa.gov/data/msl/MSLNAV_0XXX/DATA/*

# Table A2. Results

| MSL SOL | $L_S$ [deg] | Type | LTST | Sun Azim [deg] | Sun Elev [deg] | $r_{eff}$ [μm] | $\tau$ | DHG, $g_1$ | DHG, $g_2$ | DHG, $\alpha$ | $\epsilon'_0$ | $\chi^2_{SUN}$ | $\chi^2_{DHG}$ | $\chi^2_{MOP}$ |
|---|---|---|---|---|---|---|---|---|---|---|---|---|---|---|
| 2034 | 165.94 | SUN | 14.97 | 282.96 | 44.295 | $1.16^{+0.25}_{-0.08}$ | $0.80^{+0.20}_{-0.06}$ | - | - | - | - | 4.358 | - | - |
| 2080 | 192.53 | BKWD | 16.94 | 265.82 | 16.22 | - | $2.15^{+0.20}_{-0.16}$ | $0.74^{+0.19}_{-0.08}$ | $0.08^{+0.45}_{-0.78}$ | $0.90^{+0.08}_{-0.40}$ | $2.7^{+0.5}_{-0.6}$ | - | 0.379 | 0.455 |
| 2084 | 194.90 | SUN | 15.30 | 266.34 | 40.94 | $4.14^{+0.60}_{-1.09}$ | $8.46^{+0.39}_{-0.18}$ | - | - | - | - | 1.913 | - | - |
| 2086 | 196.15 | BKWD | 16.94 | 264.26 | 16.33 | - | $9.29^{+0.16}_{-2.40}$ | $0.80^{+0.10}_{-0.11}$ | $-0.80^{+0.23}_{-0.05}$ | $0.67^{+0.01}_{-0.11}$ | $3.0^{+1.0}_{-1.3}$ | - | 2.010 | 0.384 |
| 2089 | 197.82 | SUN | 11.19 | 102.38 | 77.53 | $3.34^{+0.47}_{-0.55}$ | $9.22^{+0.23}_{-0.17}$ | - | - | - | - | 0.419 | - | - |
| 2090 | 198.44 | SUN | 11.62 | 116.14 | 83.66 | $3.40^{+0.47}_{-0.51}$ | $8.60^{+0.21}_{-0.16}$ | - | - | - | - | 1.156 | - | - |
| 2092 | 199.67 | SUN | 11.70 | 126.46 | 84.40 | $3.30^{+0.47}_{-0.46}$ | $7.68^{+0.18}_{-0.17}$ | - | - | - | - | 2.257 | - | - |
| 2093 | 200.29 | SUN | 12.18 | 216.19 | 85.59 | $3.38^{+0.48}_{-0.51}$ | $8.12^{+0.20}_{-0.16}$ | - | - | - | - | 0.723 | - | - |
| 2096 | 202.24 | BKWD | 16.60 | 261.85 | 21.47 | - | $9.03^{+0.17}_{-2.89}$ | $0.78^{+0.18}_{-0.18}$ | $-0.78^{+1.39}_{-0.08}$ | $0.91^{+0.09}_{-0.41}$ | $4.3^{+0.1}_{-2.2}$ | - | 0.513 | 0.348 |
| 2097 | 202.76 | SUN | 12.85 | 249.00 | 76.73 | $3.48^{+0.49}_{-0.50}$ | $8.00^{+0.19}_{-0.16}$ | - | - | - | - | 0.661 | - | - |
| 2099 | 203.98 | SUN | 12.34 | 224.76 | 82.94 | $3.40^{+0.50}_{-0.49}$ | $7.72^{+0.19}_{-0.16}$ | - | - | - | - | 0.531 | - | - |
| 2100 | 204.61 | SUN | 12.88 | 247.15 | 76.00 | $3.42^{+0.50}_{-0.49}$ | $7.86^{+0.18}_{-0.16}$ | - | - | - | - | 0.780 | - | - |
| 2101 | 205.21 | SUN | 12.02 | 183.19 | 84.49 | $3.48^{+0.51}_{-0.48}$ | $7.42^{+0.20}_{-0.17}$ | - | - | - | - | 1.151 | - | - |
| 2102 | 205.87 | SUN | 13.75 | 255.62 | 63.29 | $3.58^{+0.54}_{-0.50}$ | $6.90^{+0.17}_{-0.14}$ | - | - | - | - | 1.070 | - | - |
| 2102 | 205.96 | BKWD | 16.99 | 260.13 | 15.71 | - | $8.67^{+0.04}_{-2.73}$ | $0.76^{+0.17}_{-0.20}$ | $-0.76^{+1.06}_{-0.08}$ | $0.89^{+0.10}_{-0.32}$ | $4.4^{+0.3}_{-2.3}$ | - | 0.829 | 0.282 |
| 2103 | 206.43 | SUN | 11.52 | 130.81 | 80.64 | $3.52^{+0.51}_{-0.48}$ | $7.44^{+0.19}_{-0.16}$ | - | - | - | - | 1.391 | - | - |
| 2106 | 208.31 | SUN | 11.89 | 166.10 | 83.04 | $3.48^{+0.53}_{-0.46}$ | $7.40^{+0.18}_{-0.17}$ | - | - | - | - | 1.773 | - | - |
| 2107 | 208.93 | SUN | 11.95 | 173.74 | 82.94 | $3.54^{+0.52}_{-0.48}$ | $7.46^{+0.19}_{-0.16}$ | - | - | - | - | 1.739 | - | - |
| 2109 | 210.18 | SUN | 11.68 | 148.64 | 81.16 | $3.46^{+0.55}_{-0.48}$ | $6.90^{+0.17}_{-0.17}$ | - | - | - | - | 1.741 | - | - |
| 2110 | 210.81 | SUN | 12.02 | 182.02 | 82.17 | $3.42^{+0.54}_{-0.46}$ | $6.90^{+0.19}_{-0.17}$ | - | - | - | - | 1.626 | - | - |
| 2110 | 210.93 | BKWD | 16.69 | 258.23 | 20.13 | - | $8.82^{+0.19}_{-2.87}$ | $0.74^{+0.25}_{-0.20}$ | $-0.74^{+1.29}_{-0.10}$ | $0.89^{+0.11}_{-0.39}$ | $4.0^{+0.1}_{-1.9}$ | - | 0.483 | 0.280 |
| 2111 | 211.44 | SUN | 12.01 | 181.32 | 81.95 | $3.52^{+0.53}_{-0.48}$ | $6.80^{+0.18}_{-0.16}$ | - | - | - | - | 1.894 | - | - |
| 2113 | 212.69 | SUN | 12.09 | 189.14 | 81.49 | $3.24^{+0.54}_{-0.45}$ | $6.44^{+0.18}_{-0.16}$ | - | - | - | - | 2.086 | - | - |
| 2114 | 213.33 | SUN | 12.26 | 203.96 | 80.46 | $3.36^{+0.52}_{-0.47}$ | $6.54^{+0.18}_{-0.16}$ | - | - | - | - | 1.841 | - | - |
| 2116 | 214.56 | SUN | 11.30 | 132.07 | 76.14 | $3.48^{+0.56}_{-0.47}$ | $6.50^{+0.17}_{-0.17}$ | - | - | - | - | 1.524 | - | - |
| 2116 | 214.63 | SAPP | 13.92 | 250.33 | 60.26 | $3.44^{+0.55}_{-0.48}$ | $6.08^{+0.15}_{-0.14}$ | - | - | - | - | 2.014 | - | - |
| 2117 | 215.34 | BKWD | 16.94 | 256.49 | 16.59 | - | $7.21^{+0.30}_{-1.82}$ | $0.73^{+0.26}_{-0.16}$ | $-0.73^{+1.32}_{-0.09}$ | $0.92^{+0.08}_{-0.42}$ | $2.5^{+1.6}_{-0.4}$ | - | 0.317 | 0.322 |
| 2118 | 215.85 | SUN | 12.14 | 190.70 | 80.23 | $3.42^{+0.56}_{-0.48}$ | $6.24^{+0.17}_{-0.17}$ | - | - | - | - | 1.356 | - | - |
| 2120 | 217.20 | SUN | 15.50 | 254.66 | 37.58 | $3.78^{+0.60}_{-0.58}$ | $5.48^{+0.10}_{-0.10}$ | - | - | - | - | 0.973 | - | - |
| 2121 | 217.82 | SUN | 15.11 | 253.44 | 43.05 | $3.72^{+0.61}_{-0.52}$ | $5.52^{+0.11}_{-0.12}$ | - | - | - | - | 1.304 | - | - |
| 2123 | 218.97 | SUN | 10.58 | 118.45 | 66.54 | $3.24^{+0.51}_{-0.45}$ | $5.50^{+0.16}_{-0.16}$ | - | - | - | - | 2.762 | - | - |
| 2125 | 220.41 | BKWD | 16.88 | 254.65 | 17.37 | - | $5.51^{+0.71}_{-0.94}$ | $0.88^{+0.11}_{-0.35}$ | $-0.02^{+0.57}_{-0.79}$ | $0.62^{+0.38}_{-0.12}$ | $1.6^{+2.5}_{-0.4}$ | - | 0.241 | 0.334 |
| 2126 | 220.96 | SUN | 13.54 | 241.36 | 64.66 | $3.12^{+0.58}_{-0.46}$ | $5.08^{+0.15}_{-0.16}$ | - | - | - | - | 1.544 | - | - |
| 2127 | 221.65 | SUN | 15.50 | 252.84 | 37.47 | $3.58^{+0.64}_{-0.54}$ | $4.64^{+0.10}_{-0.10}$ | - | - | - | - | 2.224 | - | - |
| 2128 | 222.26 | SUN | 14.30 | 248.10 | 54.27 | $3.02^{+0.58}_{-0.48}$ | $4.66^{+0.14}_{-0.14}$ | - | - | - | - | 1.085 | - | - |
| 2130 | 223.57 | SUN | 15.57 | 252.60 | 36.28 | $3.58^{+0.63}_{-0.55}$ | $4.48^{+0.09}_{-0.10}$ | - | - | - | - | 1.433 | - | - |
| 2131 | 224.18 | SUN | 14.43 | 247.80 | 52.29 | $3.26^{+0.62}_{-0.50}$ | $4.54^{+0.13}_{-0.14}$ | - | - | - | - | 1.745 | - | - |
| 2132 | 224.84 | SUN | 15.13 | 250.87 | 42.47 | $3.14^{+0.60}_{-0.49}$ | $4.14^{+0.11}_{-0.12}$ | - | - | - | - | 1.964 | - | - |
| 2133 | 225.52 | BKWD | 16.60 | 252.82 | 21.35 | - | $3.90^{+0.74}_{-0.25}$ | $0.78^{+0.10}_{-0.04}$ | $-0.72^{+1.53}_{-0.09}$ | $0.92^{+0.08}_{-0.42}$ | $2.1^{+2.0}_{-0.0}$ | - | 0.236 | 0.244 |
| 2134 | 226.13 | SUN | 15.37 | 251.03 | 38.99 | $3.14^{+0.61}_{-0.52}$ | $3.90^{+0.09}_{-0.12}$ | - | - | - | - | 1.895 | - | - |
| 2135 | 226.75 | SUN | 14.81 | 248.74 | 46.76 | $2.90^{+0.59}_{-0.49}$ | $3.92^{+0.12}_{-0.14}$ | - | - | - | - | 2.005 | - | - |
| 2139 | 229.39 | BKWD | 17.21 | 251.70 | 12.67 | - | $3.62^{+0.40}_{-0.36}$ | $0.72^{+0.06}_{-0.05}$ | $-0.69^{+0.93}_{-0.07}$ | $0.93^{+0.06}_{-0.22}$ | $2.0^{+2.0}_{-0.0}$ | - | 0.448 | 0.403 |
| 2141 | 230.63 | SUN | 15.32 | 249.21 | 39.47 | $2.86^{+0.64}_{-0.47}$ | $3.26^{+0.11}_{-0.11}$ | - | - | - | - | 2.060 | - | - |
| 2148 | 235.19 | BKWD | 16.44 | 249.69 | 23.55 | - | $2.72^{+0.29}_{-0.20}$ | $0.74^{+0.19}_{-0.05}$ | $-0.05^{+0.64}_{-0.68}$ | $0.94^{+0.05}_{-0.44}$ | $3.1^{+0.5}_{-1.2}$ | - | 0.253 | 0.283 |
| 2160 | 243.00 | BKWD | 16.90 | 248.09 | 17.08 | - | $2.08^{+0.24}_{-0.10}$ | $0.79^{+0.15}_{-0.10}$ | $0.18^{+0.25}_{-0.86}$ | $0.78^{+0.17}_{-0.28}$ | $2.3^{+0.1}_{-0.6}$ | - | 0.417 | 0.563 |
| 2165 | 246.24 | BKWD | 16.39 | 246.93 | 24.03 | - | $1.71^{+0.32}_{-0.02}$ | $0.97^{+0.02}_{-0.28}$ | $0.38^{+0.03}_{-1.08}$ | $0.51^{+0.43}_{-0.01}$ | $2.1^{+0.2}_{-0.6}$ | - | 0.338 | 1.019 |
| 2212 | 276.78 | BKWD | 17.03 | 245.33 | 15.13 | - | $1.29^{+0.08}_{-0.07}$ | $0.90^{+0.09}_{-0.09}$ | $0.20^{+0.10}_{-0.12}$ | $0.63^{+0.11}_{-0.13}$ | $1.8^{+0.1}_{-0.3}$ | - | 0.714 | 5.623 |
| 2214 | 277.96 | SAPP | 13.27 | 220.07 | 62.84 | $1.14^{+1.08}_{-0.10}$ | $1.14^{+1.31}_{-0.16}$ | - | - | - | - | 3.243 | - | - |
| 2223 | 283.81 | BKWD | 17.08 | 245.96 | 14.51 | - | $1.25^{+0.08}_{-0.06}$ | $0.88^{+0.09}_{-0.10}$ | $0.16^{+0.14}_{-0.15}$ | $0.67^{+0.11}_{-0.16}$ | $1.8^{+0.1}_{-0.2}$ | - | 0.711 | 4.590 |

Note: MSL mission sol number, solar longitude (in degrees), type of observation, local true solar time (LTST, in decimal format), solar azimuth and elevation angles (in degrees) with respect to North and local horizon, dust particle size distribution effective radius (in microns), column dust 650-nm optical depth, Double Henyey-Greenstein (DHG) analytical single scattering phase function parameters ($g_1$, $g_2$, $\alpha$) (only backscatter scenarios), aspect ratio parameter for a modified lognormal aspect ratio distribution for a mixture of prolate/oblate spheroids (only backscatter scenarios), and $\chi^2$ values for Sun-pointing, DHG and MOPSMAP backward scattering region model-to-observation comparisons, respectively.